\documentclass[aps,prb,twocolumn]{revtex4}
\usepackage{amssymb}
\usepackage{graphicx}
\usepackage{hyperref}

\begin{document}
\title{Low-dimensional magnetism in the mixed Cr valence spin-chain compound Sr$_4$Cr$_3$O$_9$}
\author{Yogesh Singh and D. C. Johnston}
\affiliation{Ames Laboratory and Department of Physics and Astronomy, Iowa State University, Ames, IA 50011, USA}
\date{\today}

\begin{abstract}
Sr$_4$Cr$_3$O$_9$ is the $n$~=~2 member of a family of quasi-one-dimensional compounds $A_{n+2}T_{n+1}$O$_{3n+3}$ ($A$~=~Ca, Sr, or Ba, $T$~=~transition metal, and $n$~=~1$–-\infty$) having a crystal structure which consists of chains of $n T$O$_6$ octahedra alternated by one $T$O$_6$ trigonal prism running along the $c$-axis.  The chains are arranged on a triangular lattice in the $ab$-plane.  We report the synthesis, structure, magnetization $M$ versus magnetic field $H$, magnetic susceptibility $\chi$ versus temperature $T$ and specific heat $C$ versus $T$ measurements on sintered and arc-melted polycrystalline samples of Sr$_4$Cr$_3$O$_9$.  The $\chi$ data have a $T$ depdendence which is typical of low-dimensional magnetic systems with dominant antiferromagnetic (AF) exchange interactions.  Specifically, $\chi(T)$ shows a broad maximum at $T_{\rm max} \approx$~200~K for the sintered pellet and $T_{\rm max}\approx$~265 K for the arc-melted sample, indicating the onset of short-range magnetic order.  Below $T$~=~15~K we observe 2 anomalies in the $\chi(T)$ data for both samples suggesting the onset of long-range magnetic ordering.  The corresponding anomalies in the $C(T)$ data however are weak indicating that only a small amount of the expected magnetic entropy is recovered at the magnetic transitions and that strong short-range AF order exists above these temperatures. 

\end{abstract}
\pacs{75.40.Cx, 75.50.Ee, 65.40.Ba, 75.30.Et}

\maketitle

\section{Introduction}
\label{sec:INTRO}
There has been a lot of interest recently in the magnetism of the quasi-one-dimensional (Q1D) oxides $A_{n+2}T_{n+1}$O$_{3n+3}$ ($A$~=~Ca, Sr, and Ba, $T$~=~Mn--Zn, Rh, Ir, and Pt and $n$~=~$1–-\infty$). \cite{onnerud1996,nguyen1994,nguyen1995,kageyama1998,Niitaka1999,kageyama1997,aasland1997,Flahaut2003,cao2007,niazi2002,sampath2002,rayaprol2003,rayaprol2004,mohapatra2007} 
The structure of these materials consists of $T$O$_3$ chains built of repeating units of $n$ face-sharing $T$O$_6$ octahedra and one face-sharing $T$O$_6$ trigonal prism running along the $c$-axis.  These chains form a two-dimensional (2D) triangular lattice in the $ab$-plane.\cite{onnerud1996,nguyen1994,Boulahya1999,Boulahya1999a,whangbo2001,fjelivag1996}  
The magnetic interactions in these materials are thought to be ferromagnetic along the $T$O$_3$ chains and antiferromagnetic (AF) between chains.\cite{nguyen1995,kageyama1998,Niitaka1999,aasland1997,kageyama1997}  Most of the studies have been on $n$~=~1 members of the series and interesting magnetic behavior has been reported.  The magnetic susceptibility $\chi$ for the compound Sr$_3$NiIrO$_6$ shows an abrupt drop below $T$~=~15~K and a singlet ground state has been suggested.\cite{nguyen1995}  A similar magnetic behavior has been observed in the isostructural materials Ca$_3$CoRhO$_6$ and Ca$_3$CoIrO$_6$.\cite{Niitaka1999}  The magnetic behavior of Sr$_3$CuPtO$_6$ has been modeled by an $S$~=~1/2 Heisenberg linear chain antiferromagnetic model.\cite{nguyen1994}  The compound Ca$_3$Co$_2$O$_6$ is the most studied member of this family of spin chain compounds and it has been shown to exhibit unusual magnetic properties.  The AF coupling between the chains and their triangular arrangement leads to frustration and a partially disordered antiferromagnetic state has been suggested where the Co spins within the CoO$_3$ chains couple ferromagnetically and the chains act as single large Ising spins.\cite{aasland1997}  Below $T\approx$~25~K two out of every three chains order ferromagnetically while one of the chains orders antiferromagnetically with respect to the other two chains.\cite{aasland1997,kageyama1997}  At lower temperatures further complex magnetic behavior has been reported and a partially disordered antiferromagnetic state has been suggested.\cite{kageyama1997,takeshita2006}

Only recently have the $n\geq 2$ members of this family been investigated for their magnetic, electrical and thermal properties.\cite{Takami2004,Sugiyama2005,Sugiyama2006,cao2007,Cao2007a}    
   
Very few compounds containing Cr in these structures have been studied.  There is only one report of the synthesis and magnetic studies on the $n$~=~1 materials Sr$_3$$T$CrO$_6$ ($T$~=~In, Sc, Y, Ho--Yb, and Lu) containing Cr$^{+3}$.\cite{smith2000}  The magnetic behavior of these materials was consistent with Curie-Weiss law at high temperatures with antiferromagnetic interactions and a broad maximum in $\chi(T)$ is observed for some of the compounds with the temperature of the maximum being in the range $T_{\rm max}$~=~15--25~K\@.\cite{smith2000}  None of these materials were reported to undergo long range magnetic order down to 2~K\@.
 
\begin{figure}[t]
\includegraphics[width=3.in,]{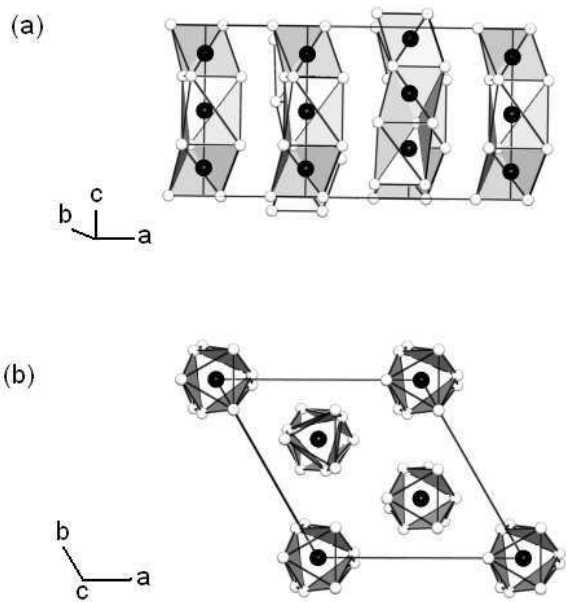}
\caption{The crystal structure of Sr$_4$Cr$_3$O$_9$.  The Cr atoms are shown as filled black spheres while the oxygen atoms are shown as open white spheres.  The CrO$_6$ polyhedra are shown in grey.  The Sr atoms have not been shown for clarity.  (a) A segment of the crystal structure of Sr$_4$Cr$_3$O$_9$ showing the CrO$_3$ chains running along the $c$-axis.  (b) The crystal structure viewed along the $c$-axis showing the triangular arrangement of the CrO$_3$ chains in the $ab$-plane.  
\label{Figstructure}}
\end{figure}

We have successfully synthesized the $n$~=~2 compound Sr$_4$Cr$_3$O$_9$ containing Cr in a formal valence state of $+3.33$. The compound Sr$_4$Cr$_3$O$_9$ was first prepared by Cuno \emph{et al.}\ and its single crystal structure was reported.\cite{Cuno1989}  The crystal structure is shown in Fig.~\ref{Figstructure}.  Figure~\ref{Figstructure}(a) shows the Cr chains extending along the $c$-axis and Fig.~\ref{Figstructure}(b) shows the triangular arrangement of the chains in the $ab$-plane.  There are 9 different crystallographic Cr positions.  Each unit cell consists of three inequivalent CrO$_3$ chains and each chain is made up of three different Cr ions.  Apart from the synthesis and crystal structure, an experimental study of the physical properties of this material has not been reported before.

Herein we report the magnetic and thermal properties of polycrystalline samples of Sr$_4$Cr$_3$O$_9$ prepared by solid-state synthesis and by arc-melting.  The magnetic susceptibility $\chi$ versus temperature $T$ data show behavior typical of low-dimensional magnetic systems with a broad maximum at $T_{\rm max} \approx$~200~K for the sintered pellet and $T_{\rm max}\approx$~265 K for the arc-melted sample.  Below $T$~=~15~K we observe two anomalies in the $\chi(T)$ data suggesting the onset of long-range magnetic ordering although the heat capacity $C$ data show only weak features at these temperatures.  The magnetization $M$ versus magnetic field $H$ data are linear at all temperatures.

\section{EXPERIMENTAL DETAILS}
\label{sec:EXPT}
Polycrystalline samples ($\sim$2~g) of Sr$_4$Cr$_3$O$_9$ were prepared by solid state synthesis.  The starting materials used were SrCO$_3$ (99.99\%, JMC) and Cr$_2$O$_3$ (99.995\%, MV Labs).  The starting materials were taken to keep the Sr:Cr atomic ratio 4:3, mixed thoroughly in an agate mortar and the powder was placed in an Al$_2$O$_3$ crucible and reacted at 1100~$^\circ$C for 24~hrs in a flow of industrial grade Ar containing about 1\% air.  The use of industrial grade Ar was needed to achieve the slight oxidation required to synthesize the material.  After this initial treatment a brick orange product is obtained.  This material is ground into a fine powder and pressed into a 1/2" pellet and fired at 1200~$^\circ$C in the above Ar atmosphere for 72~hrs with an intermediate grinding and pelletizing step after the first 48~hrs.  The resulting pellet is dark brown and well sintered.  Part of the sintered pellet was arc-melted in high purity Ar atmosphere on a water cooled copper hearth.  
The arc-melted ingot was then annealed at 1200~$^\circ$C in the industrial grade Ar atmosphere for 72~hrs to homogenize the sample.  
Powder X-ray diffraction patterns on both samples were obtained at room temperature using a Rigaku Geigerflex diffractometer with Cu K$\alpha$ radiation, in the 2$\theta$ range from 10 to 90$^\circ$ with a 0.02$^\circ$ step size. Intensity data were accumulated for 5~s per step.  The $\chi(T)$ and $M(H)$ were measured using a commercial Superconducting Quantum Interference Device (SQUID) magnetometer (MPMS5, Quantum Design) and the $C(T)$ was measured using a commercial Physical Property Measurement System (PPMS5, Quantum Design).  The dc four-probe resistivity was measured using the PPMS on a bar cut from the arc-melted sample.  The data showed insulating behavior at low temperatures with $\rho(300~{\rm K})$~=~22~$\Omega$~cm and $\rho(82~{\rm K})$~=~1~M$\Omega $~cm.

\section{RESULTS}
\subsection{Structure of Sr$_4$Cr$_3$O$_9$}
\label{sec:RES-structure}

\begin{figure}[t]
\includegraphics[width=3.in]{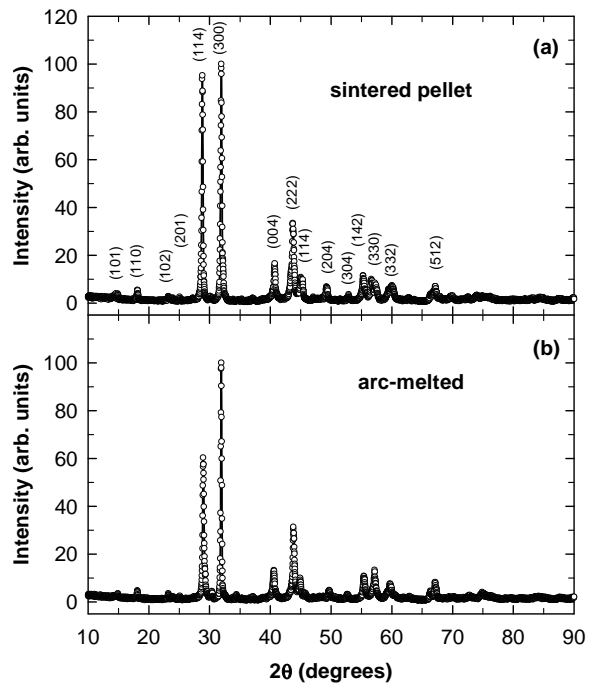}
\caption{Powder X-ray diffraction data for the sintered pellet (a) and arc-melted sample (b) of Sr$_4$Cr$_3$O$_9$.  The main peaks have been indexed using hexagonal lattice parameters to the trigonal Sr$_4$Ni$_3$O$_9$ structure.  The lines through the data points are guides to the eye.
\label{Figxrd}}
\end{figure}

Powder X-ray diffraction (XRD) patterns for the sintered pellet and annealed arc-melted samples are shown in Figs.~\ref{Figxrd}(a) and (b), respectively.  
We obtain XRD patterns in which most of the lines matched the pattern of trigonal Sr$_4$Cr$_3$O$_9$ (space group $P$3, hexagonal lattice parameters $a$~=~9.618~\AA, and $c$~=~7.874~\AA) with 9 inequivalent Cr positions, but some lines between 2$\theta$~=~20$^\circ$ and 42$^\circ$ were absent.  On the other hand, as shown in Fig.~\ref{Figxrd}(a), all the lines in the XRD pattern for our samples could be indexed using the closely related trigonal Sr$_4$Ni$_3$O$_9$ structure (Space group $P$321, hexagonal lattice parameters $a$~=~9.477~\AA, and $c$~=~7.826 ~\AA) with 5 inequivalent Ni positions.\cite{Abraham1994, Huve1998}  The overall structures of Sr$_4$Cr$_3$O$_9$ and Sr$_4$Ni$_3$O$_9$ are very similar with CrO$_3$ (NiO$_3$) chains along the $c$-axis and a triangular arrangement of the chains in the $ab$-plane.  The difference between the structures is that in Sr$_4$Cr$_3$O$_9$ some of the Cr atoms are located away from the center of the CrO$_6$ octahedra resulting in more inequivalent Cr positions in Sr$_4$Cr$_3$O$_9$ as compared to Ni positions in Sr$_4$Ni$_3$O$_9$.  

Rietveld refinement of the X-ray patterns gave the lattice parameters $a$~=~$b$~=~9.601(2) \AA\ and $c$~=~7.858(2) \AA\ for the sintered pellet and $a$~=~$b$~=~9.610(1) \AA\ and $c$~=~7.838(1) \AA\ for the annealed arc-melted sample.  These values are in reasonable agreement with the above previously reported values for Sr$_4$Cr$_3$O$_9$.\cite{Cuno1989}  It should be pointed out that only the lattice parameters, the peak profile parameters, and the atomic positions of the Sr and Cr were varied in the fits.  When the fractional occupancies and thermal parameters were varied, the fits did not converge.     

\subsection{Magnetization Measurements}
\subsubsection{Magnetic Susceptibility}
\label{sec:susceptibility}
\begin{figure}[t]
\includegraphics[width=3in]{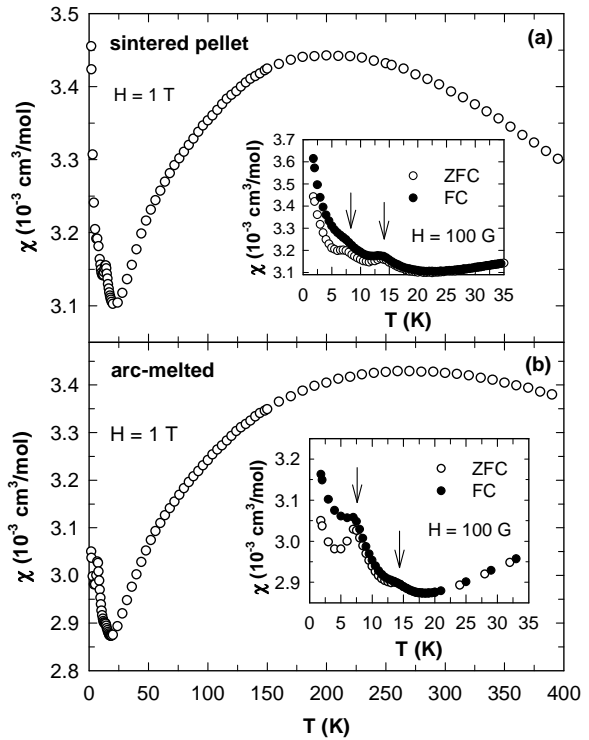}
\caption{(a) The magnetic susceptibility $\chi$ versus temperature $T$ for the sintered pellet of Sr$_4$Cr$_3$O$_9$ measured in an applied magnetic field  $H$~=~1~T\@.  The inset in (a) shows the zero-field-cooled (ZFC) and field-cooled (FC) $\chi (T)$ data between 1.8~K and 35~K for the sintered pellet measured in $H$~=~100~G\@.  The two magnetic anomalies are indicated by the arrows.  (b) $\chi$ versus $T$ for the arc-melted Sr$_4$Cr$_3$O$_9$ sample measured in an applied magnetic field of 1~T\@.  The inset in (b) shows the ZFC and FC $\chi (T)$ data between 1.8~K and 35~K for the arc-melted sample measured in $H$~=~100~G\@.  The two magnetic anomalies are indicated by the arrows. 
\label{Figsus}}
\end{figure}
\noindent
The temperature dependence of the magnetic susceptibilities $\chi \equiv M/H$ between 1.8~K and 400~K for the sintered pellet and the arc-melted annealed samples of Sr$_4$Cr$_3$O$_9$ measured in an applied magnetic field $H$~=~1~T are shown in Figs.~\ref{Figsus}(a) and (b), respectively.  The $\chi(T)$ data for both samples show a behavior typical of low-dimensional antiferromagnetic systems with a broad maximum in $\chi(T)$ at $T_{\rm max}$~=~200~K and 265~K for the sintered pellet and arc-melted annealed samples, respectively, indicating large antiferromagnetic exchange interactions ($\sim$~200~K--300~K) in this compound.  The zero-field-cooled (ZFC) and field-coolded (FC) data measured at $H$~=~100~Oe between $T$~=~1.8~K and 35~K are shown in the insets in Figs.~\ref{Figsus}(a) and (b) for the two samples, respectively.  Two anomalies, at $T$~=~13.5~K and 7.5~K, can be seen in the $\chi(T)$ data for both samples and the ZFC and FC data for both samples bifurcate below the temperature of the first anomaly at $T$~=~13.5~K\@.  These anomalies may indicate the onset of long-range magnetic ordering in this material.  The upturn in $\chi(T)$ at the lowest temperatures could be due to the presence of small amounts of paramagnetic impurities in the samples. 

Although the qualitative behavior of $\chi(T)$ for both samples is very similar, the absolute values of $\chi(T)$ for the two samples are slightly different.  This may be due to oxygen vacancies in the arc-melted sample where we had observed a small mass loss after melting.  Oxygen defficiency could lead to changes in the valance of Cr which could in turn lead to changes in the magnetic properties.  

\subsubsection{Isothermal Magnetization versus Magnetic Field}
\label{sec:MH}
\begin{figure}[t]
\includegraphics[width=3in]{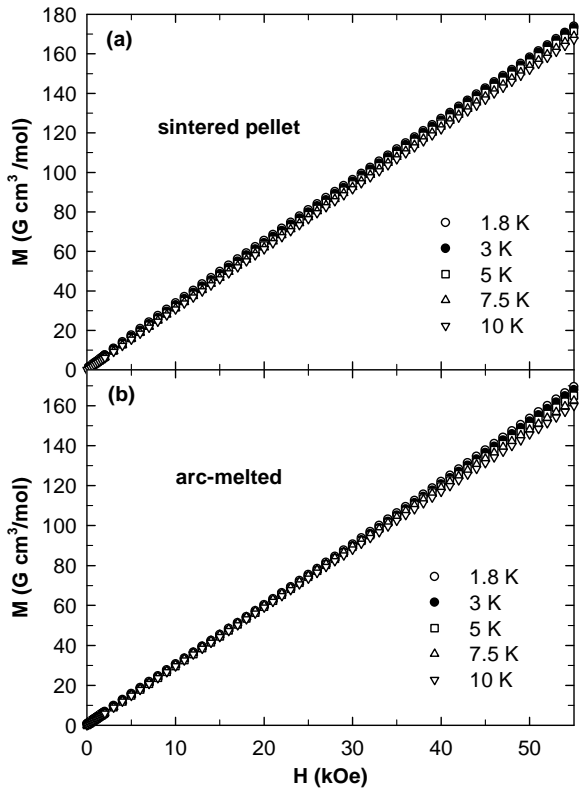}
\caption{ Isothermal magnetization $M$ versus magnetic field $H$ at various temperatures for (a) the sintered pellet and (b) for the arc-melted samples of Sr$_4$Cr$_3$O$_9$.
\label{FigMH}}
\end{figure}
\noindent
Figures~\ref{FigMH}(a) and (b) show the isothermal magnetization $M$ versus magnetic field $H$ measured at various temperatures $T$ for the sintered pellet and arc-melted samples, respectively.  The $M(H)$ data for both samples are linear at all temperatures with no sign of curvature or saturation.  The $M(H)$ data at higher temperatures (not shown) are also linear.  The fact that $M$ does not saturate even at $H$~=~6~T at 1.8~K may indicate that the upturn in $\chi(T)$ at the lowest temperature is not due to paramagnetic impurities but may be intrinsic to the material.

\subsection{Heat Capacity}
\label{sec:RES-heatcapacity}
\begin{figure}[t]
\includegraphics[width=3in]{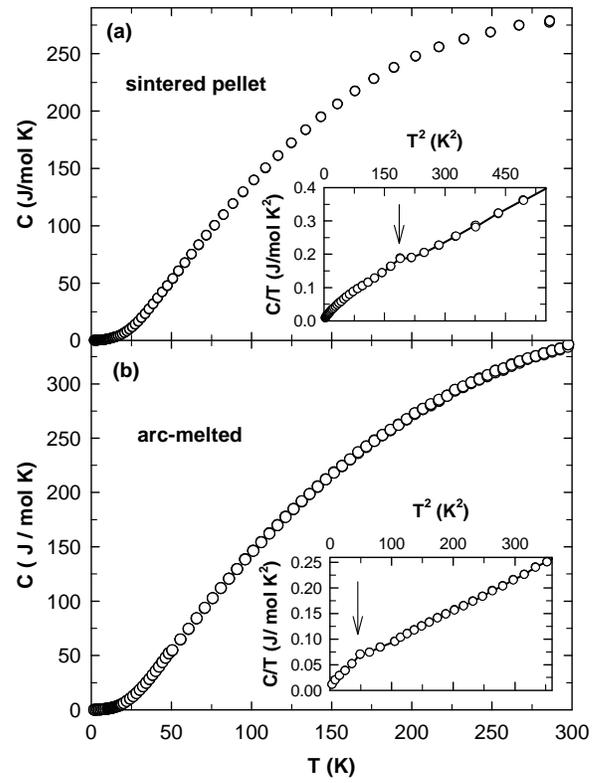}
\caption{(a) The heat capacity $C$ versus temperature $T$ for a sintered pellet of Sr$_4$Cr$_3$O$_9$ between 1.8~K and 300~K\@.  The inset in (a) shows the $C/T$ versus $T^2$ data between 1.8~K and 23.5~K\@.  (b) $C$ versus $T$ of the arc-melted annealed sample of Sr$_4$Cr$_3$O$_9$ between 1.8~K and 300~K\@.  The inset in (b) shows the the $C/T$ versus $T^2$ data between 1.8~K and 19~K\@.  The arrows in (a) and (b) indicate the positions of the magnetic transitions.
\label{FigHC}}
\end{figure}

The heat capacity $C$ versus temperature $T$ of the sintered pellet and annealed arc-melted samples of Sr$_4$Cr$_3$O$_9$ between 1.8~K and 300~K are shown in Figs.~\ref{FigHC}(a) and (b), respectively.  There is no anomaly in the $C(T)$ data for either sample at $T_{\rm max}$ where the maximum in $\chi(T)$ was observed.  The inset in Fig.~\ref{FigHC}(a) shows the $C/T$ versus $T^2$ data for the sintered pellet of Sr$_4$Cr$_3$O$_9$ between 1.8~K and 23.5~K to highlight the data in the temperature range where the two magnetic anomalies were observed in the $\chi(T)$ data in Fig.~\ref{Figsus}(a).  Only a weak and broad anomaly with a maximum at $T$~=~13~K is observed corresponding to the $T$~=~13.5~K anomaly observed in the $\chi(T)$ data.  No obvious anomaly is observed in the $C(T)$ data corresponding to the second magnetic anomaly that was seen in the $\chi(T)$ data at $T$~=~7.5~K\@.  The inset in Fig.~\ref{FigHC}(b) shows the low temperature $C/T$ versus $T^2$ data between 1.8~K and 19~K for the arc-melted annealed sample of Sr$_4$Cr$_3$O$_9$.  A single broad anomaly peaked at $T$~=~7.5~K is observed corresponding to the lower temperature anomaly observed in the $\chi(T)$ data in Fig.~\ref{Figsus}(b) for this sample.  We were unable to synthesize a non-magnetic material in this structure which could act as a reference compound for the lattice heat capacity and help in the extraction of the magnetic contribution to $C(T)$ and the magnetic entropy associated with the anomalies observed in $C(T)$.  The weak features in the $C(T)$ data could either suggest partial ordering as has been observed in other members of this family of spin-chain compounds,\cite{cao2007,Cao2007a} or may indicate that most of the magnetic entropy is recovered over a wide temperature range above the long range ordering temperatures as indicated by the broad maxima in the $\chi(T)$ data for these samples at 200--300~K\@.  This observation is common in low-dimensional magnetic systems.

\begin{figure}[t]
\includegraphics[width=3in]{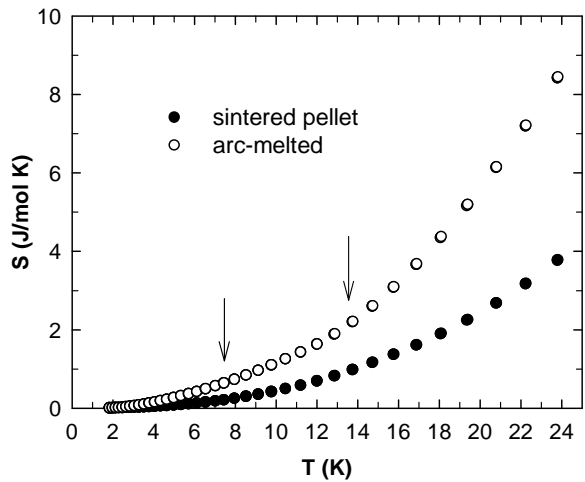}
\caption{The entropy $S$ of the sintered pellet and the arc-melted samples of Sr$_4$Cr$_3$O$_9$ versus temperature $T$ between 1.8~K and 24~K\@.  The arrows indicate the temperatures at which the anomalies in the $\chi(T)$ data were observed.    
\label{Fig-entropy}}
\end{figure}

To get an upper limit estimate of the magnetic entropy $S_{\rm mag}(T)$ recovered up to the transition at $T$~=13.5~K we have computed the total magnetic plus lattice entropy  $S(T)$~=~$\int ^ T_{1.8~{\rm K}}[C(T)/T]dT$ between $T$~=~1.8~K and 24~K as shown in Fig.~\ref{Fig-entropy} for both samples.  The arrows in the figure indicate the temperatures where anomalies were observed in the $\chi(T)$ data in Figs.~\ref{Figsus}(a) and (b).  In Sr$_4$Cr$_3$O$_9$, the 3 Cr ions have a charge of $+10$.  Assuming that the charge is distributed as two Cr$^{+3}$ (spin $s$~=~3/2) and one Cr$^{+4}$ ($s$~=~1) the expected disordered magnetic entropy would be $S_{\rm mag}$~=~R$[2\ln(4) + \ln(3)]$~=~32.2~J/mol~K\@.  At $T$~=~14~K we observe $S$~=~1.0~J/mol~K and $S$~=~2.4~J/mol~K for the sintered pellet and arc-melted samples, respectively.  These estimates also include the lattice contribution thus giving upper limit estimates of $S_{\rm mag}$ and show that only a small fraction ($< 7\%$) of the magnetic entropy is recovered at the magnetic ordering temperatures. 

\begin{figure}[t]
\includegraphics[width=3in]{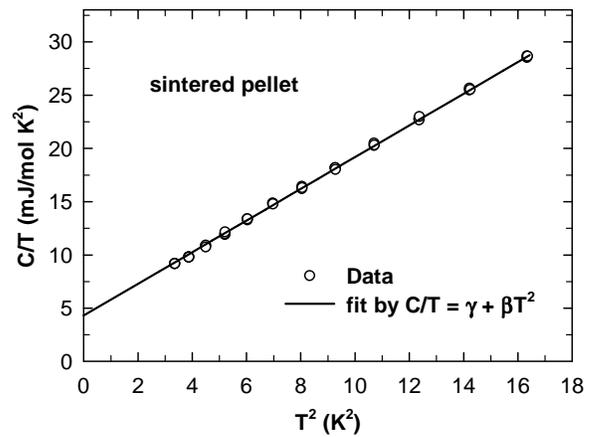}
\caption{The heat capacity of the sintered pellet of Sr$_4$Cr$_3$O$_9$ plotted as $C/T$ versus temperature squared $T^2$ between 1.8~K and 4~K\@.  The solid line through the data is a linear fit by the expression $C/T$~=~$\gamma + \beta T^2$.    
\label{FigHC-lowT}}
\end{figure}

The $C(T)/T$ data of the sintered pellet of Sr$_4$Cr$_3$O$_9$ between 1.8~K and 4~K are shown in Fig.~\ref{FigHC-lowT} and could be fitted by the expression $C/T$~=~$\gamma + \beta T^2$.  The fit shown as the solid line through the data in Fig.~\ref{FigHC-lowT} gave the values $\gamma$~=~4.3(2)~mJ/mol K$^2$ and $\beta$~=~1.49(1)~mJ/mol K$^4$.  The small but finite value of $\gamma$ in this insulating compound could arise from disorder or small amounts of impurity phases.  We note that a large $\gamma \sim 100$~mJ/mol K$^2$ has been observed for some other insulating low-dimensional magnetic systems like $L_2$RuO$_5$ ($L$~=~Pr, Nd, Sm, Gd, and Tb),\cite{cao2001} and Sr$_5$Rh$_4$O$_{12}$.\cite{Cao2007a}  From the value of $\beta$ one can obtain the Debye temperature $\theta_{\rm D}$ using the expression \cite{Kittel}
\begin{equation}
\Theta_{\rm D}~=~\bigg({12\pi^4{\rm R} p \over 5\beta}\bigg)^{1/3}~, 
\label{EqDebyetemp}
\end{equation}
\noindent
where R is the molar gas constant and $p$ is the number of atoms per formula unit (\emph{p}~=~16 for Sr$_4$Cr$_3$O$_9$).  We obtain $\Theta_{\rm D}$~=~276(2)~K for the sintered pellet of Sr$_4$Cr$_3$O$_9$.  This is an extremely low value for an oxide material.  It must be noted that a $T^3$ temperature dependence of the heat capacity is also expected for 3-dimensional antiferromagnetic magnons below the antiferromagnetic transition temperature.  Therefore, the $\beta$ obtained above has contributions from both the phonons and the AF magnons.  Thus the $\theta_{\rm D}$ estimated above from $\beta$ is a lower limit estimate.  This is consistent with the fact that even at $T$~=~300~K the $C$ for both samples (280--330~J/mol~K) is smaller than the classical Dulong-Petit value $C$~=~3pR~=~399~J/mol~K\@.

The low temperature $C(T)$ of the annealed arc-melted sample of Sr$_4$Cr$_3$O$_9$ could not be used to obtain $\theta_{\rm D}$ because of the presence of the anomaly at 7.5~K\@.  

\section{DISCUSSION}
\label{sec:DIS}
The results of our magnetic and thermal property measurements show that as expected from the structure Sr$_4$Cr$_3$O$_9$ shows low dimensional magnetic behavior with predominant antiferromagnetic interactions that are presumably along the Cr chains.  The Cr-containing $n$~=~1 materials Sr$_3$$T$CrO$_6$ ($T$~=~In, Sc, Y, Ho--Yb, and Lu) also show evidence from Curie-Weiss magnetic susceptibility $\chi(T)$ behavior at high $T$ for antiferromagnetic exchange interactions within the chains, and broad maxima in $\chi(T)$ at 15~K--20~K are also observed.\cite{smith2000}  Since the interchain distance increases with increasing $n$,\cite{Sugiyama2006} comparison of the above results for the $n$~=~1 materials Sr$_3$$T$CrO$_6$ ($T$~=~In, Sc, Y, Ho--Yb, and Lu) and our $n$~=~2 material Sr$_4$Cr$_3$O$_9$ suggests that the quasi-one-dimensional character for the Cr-containing materials increases with increasing $n$ because the interchain interaction weakens due to the increased interchain distance.  The Ni-containing compounds Sr$_3$NiIrO$_6$ and Sr$_3$NiPtO$_6$ also show predominantly antiferromagnetic interactions as evidenced by a negative Weiss temperature $\theta$ obtained from Curie-Weiss fits to $\chi(T)$ data.\cite{nguyen1994,Flahaut2003}  These Ni materials also show maxima in their $\chi(T)$ data but the temperature of the maximum is $T =~80$~K and 25~K for Sr$_3$NiIrO$_6$ and Sr$_3$NiPtO$_6$, respectively.\cite{nguyen1994,Flahaut2003}  The predominantly antiferromagnetic interactions within chains in the above materials is in contrast to the Co-containing materials of this family of compounds which all show predominantly ferromagnetic exchange interactions within the Co chains.\cite{Sugiyama2005,Sugiyama2006}

\section{CONCLUSION}
\label{sec:CON}
We have synthesized polycrystalline samples of the spin-chain compound Sr$_4$Cr$_3$O$_9$ and studied their structural, magnetic and thermal properties using sintered pellet and annealed arc-melted samples.  The $\chi(T)$ data for both samples show low-dimensional magnetic behavior with a broad maximum at $T_{\rm max}$~=~250~K for the sintered pellet and $T_{\rm max}$~=~265~K for the arc-melted sample indicating the onset of dominant short-range antiferromagnetic order.  There are two anomalies in the $\chi(T)$ data at $T_1$~=~13.5~K and $T_2$~=~7.5~K for both samples suggesting the onset of long-range magnetic order.  The ZFC and FC $\chi(T)$ data for both samples measured in a magnetic field $H$~=~100~G show a bifurcation below $T_1$~=~13.5~K\@.  The $C(T)$ data show only weak anomalies at the temperatures of the corresponding magnetic anomalies and only a fraction ($<$~7\%) of the expected magnetic entropy is recovered at these transitions indicating strong short-range order above these temperatures.  These results point to a strong low dimensional character of the magnetic exchange interactions in this material.

\begin{acknowledgments}
Work at the Ames Laboratory was supported by the Department of Energy-Basic Energy Sciences under Contract No.\ DE-AC02-07CH11358.  
\end{acknowledgments}


\begin{references}
\bibitem{onnerud1996} C. Lampe-Onnerud, M. Sigrist, and H.-C. zur Loye, J. Solid State Chem. {\bf 127}, 25 (1996).
\bibitem{nguyen1994} T. N. Nguyen, D. M. Giaquinta, and H.-C. zur Loye, Chem. Mat. {\bf 6}, 1642 (1994).
\bibitem{nguyen1995} T. N. Nguyen and H.-C. zur Loye, J. Solid State Chem. {\bf 117}, 300 (1995).
\bibitem{kageyama1998} H. Kageyama, K. Yoshimura, and K. Kosuge, J. Solid State Chem. {\bf 140}, 14Ð19 (1998).
\bibitem{Niitaka1999} S. Niitaka, H. Kageyama, M. Kato, K. Yoshimura, and Koji Kosuge, J. Solid State Chem. {\bf 146}, 137 (1999).
\bibitem{aasland1997} S. Aasland, H. Fjellvag, and B. Hauback, Solid State Commun. {\bf 101}, 187 (1997).
\bibitem{kageyama1997} H. Kageyama, K. Yoshimura, K. Kosuge, H. Mitamura, and T. Goto, J. Phys. Soc. Jpn. {\bf 66}, 1607 (1997).
\bibitem{Flahaut2003} D. Flahaut, S. Hebert, A. Maignan V. Hardy, C. Martin, M. Hervieu, M. Costes,
B. Raquet, and J.M. Broto, Eur. Phys. J. B {\bf 35}, 317 (2003).
\bibitem{cao2007} G. Cao, S. Parkin, and P. Schlottmann, Solid State Commun. {\bf 141}, 369 (2007).
\bibitem{niazi2002} A. Niazi, P. L. Paulose, E. V. Sampathkumaran, Phys. Rev. Lett. {\bf 88}, 107202 (2002).
\bibitem{sampath2002} E. V. Sampathkumaran and A. Niazi, Phys. Rev. B {\bf 65}, 180401(R) (2002). 
\bibitem{rayaprol2003} S. Rayaprol, K. Sengupta, and E. V. Sampathkumaran, Phys. Rev. B {\bf 67}, 180404(R) (2003).
\bibitem{rayaprol2004} S. Rayaprol, K. Sengupta, and E. V. Sampathkumaran, J. Solid State Chem. {\bf 177}, 3270 (2004).
\bibitem{mohapatra2007} N. Mohapatra, K. K. Iyer, S. Rayaprol, and E. V. Sampathkumaran, Phys. Rev. B {\bf 75}, 214422 (2007).
\bibitem{takeshita2006} S. Takeshita, J. Arai, T. Goko, K. Nishiyama, and K. Nagamine, J. Phys. Soc. Jpn. {\bf 75}, 034712 (2006).
\bibitem{Cuno1989} R. Cuno and H. Mueller-Buschbaum, Z. anorg. allg. Chem. {\bf 572}, 175 (1989).
\bibitem{Boulahya1999} K. Boulahya, M. Parras, and J. M. Gonzalez-Calbet, J. Solid State Chem. {\bf 142}, 419 (1999).
\bibitem{Boulahya1999a} K. Boulahya, M. Parras, and J. M. Gonzalez-Calbet, J. Solid State Chem. {\bf 145}, 116 (1999).
\bibitem{whangbo2001} M.-H. Whangbo, H.-J. Koo, K.-S. Lee, O. Gourdon, M. Evain, S. Jobic, and R. Brec, J. Solid State Chem. {\bf 160}, 239 (2001).
\bibitem{fjelivag1996} H. Fjellvag, E. Gulbrandsen, S. Aasland, A. Olsen, and B. C. Hauback, J. Solid State Chem. {\bf 124}, 190 (1996).
\bibitem{Takami2004} T. Takami, H. Ikuta, and U. Mizutani, Jpn. J. Appl. Phys. {\bf 43}, 8208 (2004).  
\bibitem{Sugiyama2005} J. Sugiyama, H. Nozaki, J. H. Brewer, E. J. Ansaldo, T. Takami, H. Ikuta, and U. Mizutani, 
Phys. Rev. B {\bf 72}, 064418 (2005).
\bibitem{Sugiyama2006} J. Sugiyama, H. Nozaki, Y. Ikedo, K. Mukai, D. Andreica, A. Amato, J. H. Brewer, E. J. Ansaldo, G. D. Morris, T. Takami, and H. Ikuta, Phys. Rev. Lett. {\bf 96}, 197206 (2006).
\bibitem{smith2000} M. D. Smith and H.-C. zur Loye, Chem. Mater. {\bf 12}, 2404 (2000).
\bibitem{Abraham1994} F. Abraham, S. Minaud, and C. Renard, J. Mater. Chem. {\bf4}, 1763 (1994).
\bibitem{Huve1998} M. Huve, C. Renard, F. Abraham, G. Van Tendeloo, and S. Amelinckx, J. Solid State Chem. {\bf 135}, 1 (1998).
\bibitem{Cao2007a} G. Cao, V. Durairaj, S. Chikara, S. Parkin, and P. Schlottmann, Phys. Rev. B {\bf 75}, 134402 (2007).
\bibitem{cao2001} G. Cao, S. McCall, Z. X. Zhou, C. S. Alexander, J. E. Crow, R. P. Guertin, and C. H. Mielke, Phys. Rev B {\bf 63}, 144427 (2001).
\bibitem{Kittel} C. Kittel, \emph{Introduction to Solid State Physics}, 4th edition (John Wiley and Sons, Inc., New York, 1966).
\end{references}
\end{document}